%% file: main.tex
\documentclass[conference]{IEEEtran}
\IEEEoverridecommandlockouts
\usepackage{cite}
\usepackage{amsmath,amssymb,amsfonts}
\usepackage{algorithmic}
\usepackage{graphicx}
\usepackage{textcomp}
\usepackage{xcolor}
\usepackage{subfig}
\usepackage{subcaption}
\usepackage{hhline}
\usepackage{tabularx}
\usepackage{arabtex}
\usepackage{caption}%
\usepackage{utf8}
\setcode{utf8}
\usepackage{color,soul}
\usepackage{multirow}
\usepackage{mathrsfs}%
\usepackage{manyfoot}%
\usepackage{booktabs}
\usepackage{hyperref}
\usepackage{algorithm}%
\usepackage{listings}%

\def\BibTeX{{\rm B\kern-.05em{\sc i\kern-.025em b}\kern-.08em
    T\kern-.1667em\lower.7ex\hbox{E}\kern-.125emX}}
\begin{document}

\title{Speak the Art: A Direct Speech to Image Generation Framework}




\author{
\textbf{Mariam Saeed\textsuperscript{1}},
\textbf{Manar Amr\textsuperscript{1}},
\textbf{Farida Adel\textsuperscript{1}},
\textbf{Nada Hassan\textsuperscript{1}}, \\
\textbf{Nour Walid\textsuperscript{1}},
\textbf{Eman Mohamed\textsuperscript{1}},
\textbf{Mohamed Hussein\textsuperscript{2}},
\textbf{Marwan Torki\textsuperscript{1}}
\\
\\
\small
es-[mariamzaho4, ManarAmr2023, Faridaadel2023, Nada.Hassan2023, nourwalid70, Eman.abdo2023]@alexu.edu.eg\\
mehussein@isi.edu, mtorki@alexu.edu.eg\\
\textsuperscript{1}Alexandria University, Alexandria, Egypt \\
\textsuperscript{2}University of Southern California, CA, USA
}

\maketitle

\begin{abstract}
Direct speech-to-image generation has recently shown promising results. However, compared to text-to-image generation, there is still a large gap to enclose. Current approaches use two stages to tackle this task: speech encoding network and image generative adversarial network (GAN). The speech encoding networks in these approaches produce embeddings that do not capture sufficient linguistic information to semantically represent the input speech. GANs suffer from issues such as non-convergence, mode collapse, and diminished gradient, which result in unstable model parameters, limited sample diversity, and ineffective generator learning, respectively. To address these weaknesses, we introduce a framework called \textbf{Speak the Art (STA)} which consists of a speech encoding network and a VQ-Diffusion network conditioned on speech embeddings. To improve speech embeddings, the speech encoding network is supervised by a large pre-trained image-text model during training. Replacing GANs with diffusion leads to more stable training and the generation of diverse images. Additionally, we investigate the feasibility of extending our framework to be multilingual. As a proof of concept, we trained our framework with two languages: English and Arabic. Finally, we show that our results surpass state-of-the-art models by a large margin.
\end{abstract}

\input{sec/1_intro}
 \input{sec/2_background}

 \input{sec/3_method}
 \input{sec/4_experiments}

\input{sec/5_results}
 \input{sec/6_ablation_study}

 \input{sec/7_conclution}

\bibliographystyle{plain} 
\bibliography{sn-bibliography}

\end{document}

%% file: sec/1_intro.tex
\section{Introduction}
\label{sec:intro}
Nowadays, we live in the era of AI and machine learning where many tasks can be automated to facilitate our daily lives. Many efforts have been carried out to enhance the performance of the text-to-image generation task \cite{6,7,8,9,10,11,12}. These efforts have reached pleasing results and succeeded in capturing the semantic information of text when generating images. It is worth mentioning that many of the world’s languages are unwritten languages that have no commonly used written form. Such languages cannot benefit from the great progress in the task of text-to-image generation. 

Text form carries the semantic information only and is often more formal and structured. This may differ from the way people naturally converse. Generally, using speech rather than text, as an input for such models, is easier because people communicate using their native spoken languages. Speech signals contain far more information such as phonetic information, prosodic information, speaker identity, environmental sounds, background noise, language variation and disfluencies. Some of this information may help to achieve better results but other information may not be related to the described image, such as speaker identity, which can mislead the model’s training process. Moreover, speech signals are continuous, and have no boundaries between words. These differences make the task of speech-to-image generation lagging behind the task of text-to-image generation by a large margin.

There are two common approaches to solve the problem of speech-to-image generation: indirect and direct approaches. The indirect approach consists of a speech-to-text translation step followed by a text-to-image generation step which requires a pipeline of an ASR module and text-to-image module which consists of text encoder and generative model. The cascade of separately learned interconnected models, each responsible for specific tasks or processing stages within the system, may introduce compound errors, propagated errors, and information bottlenecks. Trying to simplify this pipeline, the direct approach was introduced, eliminating the intermediate-level text.

We introduce a two-stage framework for the task of direct speech-to-image generation. Our work is inspired by the prior work of Li et al. \cite{1} and S2IGAN \cite{2}. Our STA builds on these approaches to synthesize high-quality images consistent with their spoken descriptions. In the first stage of STA, we use a speech encoding network to encode speech semantics into an embedding vector. For the second stage, we use a diffusion model to generate images based on speech embeddings. Then, we try to extend our framework to be multilingual by replacing the speech encoding network with a multilingual one. Finally, we provide a detailed ablation study to show the effectiveness of each component in our framework and report the results on the most commonly used datasets, CUB \cite{31}, Oxford-102 \cite{39} and Flickr8k \cite{20}. The results show that our method sets a new state-of-the-art for the speech-to-image task and provides a great enhancement for unwritten or low-resourced languages.

The main contributions of this work can be summarized as follows:
\begin{itemize}
    \item Our work contributes  to enhancing the performance of direct speech-to-image generation to reduce the gap between speech-to-image and text-to-image generation.
    \item To the best of our knowledge, we are the first direct speech-to-image framework to utilize VQ-Diffusion for image synthesis.
    \item We show how the model can be extended to be multilingual. Additionally, we are the first to use Arabic datasets for this task.
    \item Experiments on two of the most popular benchmark datasets show that our framework is the state of the art in the task of speech-to-image generation.
\end{itemize}


%% file: sec/2_background.tex
\section{Background and Related Work}
\subsection{Image Generation}
The remarkable development of GANs has opened the door for various high-dimensional data generation tasks since they were introduced in \cite{3}. GANs consist of generators and discriminators. The generator is used to generate fake data similar to the real data. The discriminator’s goal is to differentiate between fake and real images. Conditional GANs presented in \cite{4} are used to generate images in the text-to-image task based on a condition vector. Conditional GANs can be used the same way in the speech-to-image task by replacing the text encoder with a speech encoder, then the vector representations extracted from the speech can be given as a condition to GAN. Another popular type of generative models is Variational Autoencoders (VAE) \cite{33} that consist of encoders and decoders. Similar to vanilla autoencoders, the encoder in a VAE encodes input data into hidden states, and the decoder tries to reconstruct the input back from the encoding. Different from vanilla autoencoders, VAE's loss includes a regularization term that encourages the embedding space to follow a certain prior distribution. VQ-VAE is another type of VAE introduced in \cite{34}. VQ-VAE uses vector quantization to obtain a discrete latent representation. VQ-VAE differs from VAE in two key ways: the encoder network outputs discrete, rather than continuous, codes; and the prior is learned rather than being fixed.

Recently, diffusion models have gained great popularity as they balance training stability and quality. The idea of diffusion models can be simply defined through two stages. In the first stage, the model works on corrupting the data by adding noise progressively. In the second stage, the corruption process is reversed to generate back the original data by denoising the corrupted data. It has been shown in \cite{5} that the diffusion models are superior to GAN-based models with some limitations. 

\subsection{Text-to-Image}
Text-to-image models were inspired by the development of GANs. The first model was introduced in \cite{4}, where they used a GAN to generate images conditioned on vector representations extracted from the text description. This model was further improved by using a stacked structure in StackGAN \cite{6}. StackGANs were proposed to generate higher-resolution images through multiple steps. It has been shown in \cite{7} that the architecture can be further improved by adding a word-level mechanism which led to the development of the AttnGAN model. Later, MirrorGAN \cite{8} added an image-to-text module, resulting in an encoder-decoder architecture. MirrorGAN ensures that the synthesized pictures are semantically consistent with the text description by recovering the text description from the images. A siamese structured GAN called SEGAN was introduced in \cite{9} to learn consistent high-level semantics.

Keeping up with the development of diffusion models, VQ-Diffusion \cite{11} emerged and outperformed the previous text-to-image generation models. The VQ-Diffusion model overcomes the unidirectional bias and the accumulated prediction error issues which were considered as weaknesses in the previous models. The VQ-Diffusion model can generate more complex scenes, which surpasses previous GAN-based text-to-image methods. Recently, VQ-Diffusion was followed by the improved VQ-Diffusion \cite{12} which improves the samples' quality and their consistency with the input text.

\subsection{Speech-to-Image}
The correlation between images and audio has encouraged many studies in various generation tasks. Chen et al. \cite{13} were the first to approach this cross-modality generation problem. Classifier-based feature extractors and conditional GANs were used to define a sound-to-image network. This network aims to generate images of the instruments from the sound of music. Hao et al. \cite{14} extended the network for mutual visual-audio generation. 

Our main concern in this paper is to generate images conditioned on the speech description. This task was first approached by Li et al. \cite{1}. They trained the speech encoder via teacher-student learning to transfer the knowledge from a pre-trained image encoder into the speech encoder. Then, the speech encoder generates embedding features to be fed into the conditional GAN. Later, Wang et al. \cite{2} followed up and introduced a different strategy in the S2IGAN model. They improved the generative model by introducing a new relation-supervised densely-stacked generative model to synthesize images that are semantically consistent with the learned embeddings. Zhang et al. \cite{15} extended S2IGAN and proposed Fusion-S2iGan, which only uses a generator/discriminator pair to project a speech signal into a high-quality picture directly.
\subsection{Multilingual Models}

Multilingual model training has two approaches. One approach was taken by \cite{17}, where they tripled their encoder to work with three languages. Accordingly, an encoder is added for each language. However, this approach would require different strategy during inference because the language of every speech has to be mapped to the corresponding encoder. Another approach could be seen in \cite{18}, where they built a single mHuBERT model for three languages by combining their data without applying any language-dependent weights or sampling. The same approach was applied in M-SpeechCLIP \cite{19} where they showed that the CLIP semantic embedding space can be used to represent not only multiple modalities but also multiple languages effectively. So, they trained a single model to encode speech in multiple languages into the same embedding space.  M-SpeechCLIP was proven to achieve high performance for non-English languages as well as English.

%% file: sec/3_method.tex
\section{Method}
\begin{figure}[!ht]
  \centering
  \includegraphics[width=1\linewidth]{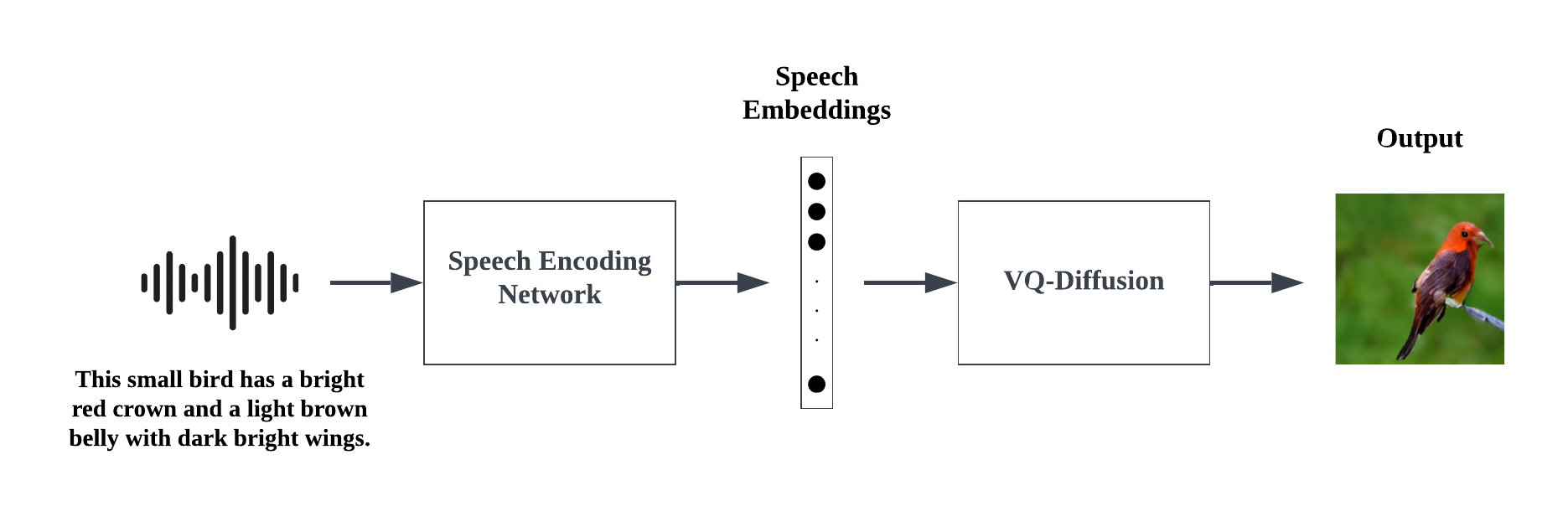}
  \caption{The whole framework of STA. It consists of two stages: a speech encoding network to extract embeddings and a VQ-Diffusion model to generate high-quality images. Text is only used for illustration.}
  \label{fig:full_model}
\end{figure}

Given a spoken description of an image, our goal is to generate an image semantically aligned with the content of the input speech signal. Our model, as shown in Figure \ref{fig:full_model}, consists of two stages, a speech encoding network to create the speech embeddings and a VQ-Diffusion model to generate the images conditioned on these speech embeddings.

\subsection{First Stage: Speech Encoding Network}
\subsubsection{Training}
\begin{figure}[!ht]
  \centering
  \includegraphics[width=0.65\linewidth]{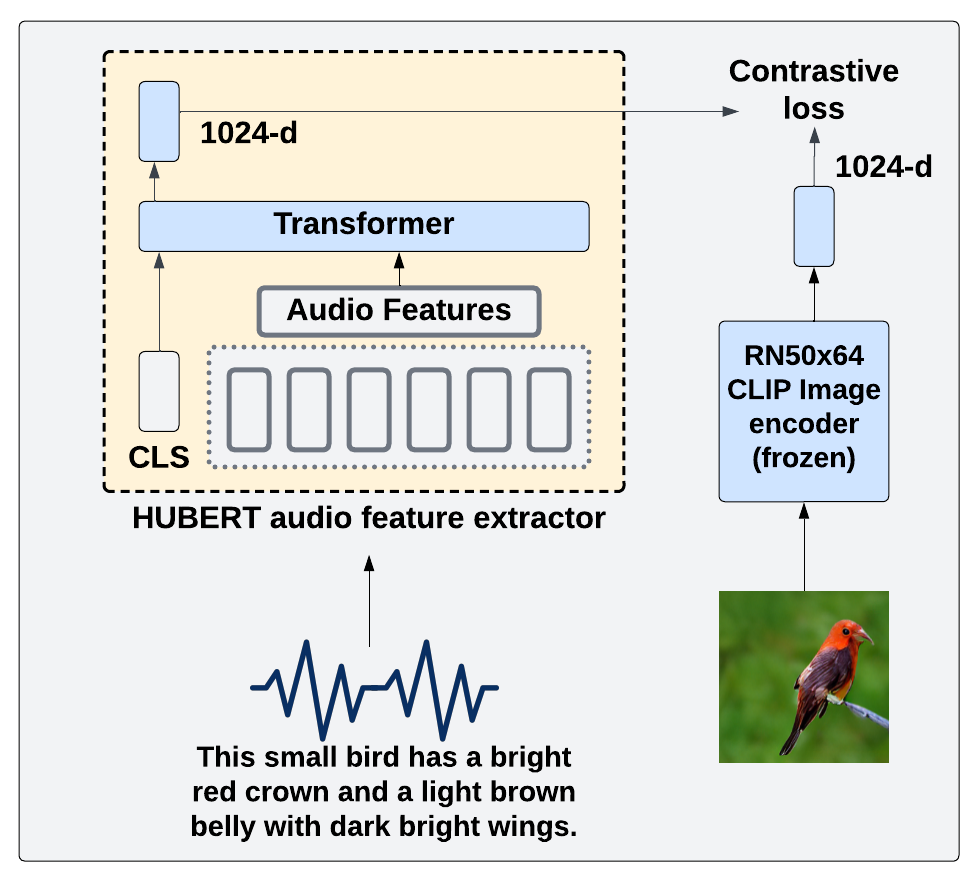}
  \caption{The architecture of the speech encoding network.}
  \label{fig:speech_net}
\end{figure}

The first stage in our framework is a speech encoding network that is based on parallel SpeechCLIP \cite{16}. We train the speech encoding network from scratch to produce $1024-d$ embeddings. During training, the input of this stage is pairs of images and their corresponding spoken captions. As demonstrated in Figure \ref{fig:speech_net}, the first stage consists of two parts:
\begin{itemize}
    \item Contrastive Language-Image Pre-training (CLIP) \cite{27} to encode the image.
    \item Pretrained Hidden-unit BERT (HuBERT) \cite{26} to produce speech embeddings.
\end{itemize}

CLIP is used to create image embeddings supervised by its corresponding text caption to capture its semantic meanings. CLIP is used as a frozen module that is pre-trained on large-scale datasets so we do not need to include text as input during training. We use RN50x64 CLIP to extend the large parallel SpeechCLIP \cite{16} producing an image embedding vector $x_i$ of dimension $1024$.

HuBERT extends the BERT framework to the audio domain by leveraging masked acoustic modeling. HuBERT consists of two essential components:
\begin{itemize}
    \item  Acoustic Feature Extraction (CNN): to obtain features that capture the spectral information of the input audio signal.
    \item Transformer encoder, which consists of stacked transformer layers. Each transformer layer includes self-attention mechanisms and feed-forward neural networks \cite{28}.
\end{itemize}

HuBERT’s output is a sequence of frame-level speech features. A learnable CLS token is added at the beginning of each sequence. The sequence is passed through a transformer encoder to obtain a speech embedding vector $y_i$ with a $1024-d$ for the following stage.

Speech embeddings are used to compute the cosine similarity with image embeddings in a mini-batch for calculating the contrastive loss \cite{27,36}. Contrastive loss encourages higher similarity scores for real speech and image pairs and lower scores for incorrect pairs. This effectively aligns speech and visual information.
\subsubsection{Inference}
During inference, The CLIP image encoder is not needed. The speech encoding network takes only spoken captions and outputs their semantically consistent speech embeddings by passing them through the HuBERT audio feature extractor. The extracted speech embeddings are then used in the second stage to generate the output image.

\subsection{Second Stage: VQ-Diffusion Model} 
Inspired by the great success of \cite{11,12}, we propose a modified version of VQ-Diffusion as our second stage to generate output images. The architecture in prior work \cite{11} consists of a CLIP encoder to encode the text, a vector quantized variational autoencoder, VQ-VAE \cite{34}, to encode and decode discrete tokens and a diffusion decoder to produce these tokens. However, we do not use a CLIP encoder. In addition, we modify the diffusion decoder to take speech embeddings as input instead of text to guide the generation of the images. As shown in Figure \ref{fig:diffusion}(b), the second stage mainly has two components: a pre-trained VQ-VAE and a diffusion image decoder.
\subsubsection{Training}
The input in the training stage is a pair of the image and its corresponding spoken caption encoded as an embedding vector. First, the input image is encoded, using VQ-VAE, into discrete tokens that represent the spatial features of the image. The image encoding process compresses the image to reduce computation overhead. The image is represented by a token vector $k \in \mathbb{Z}^N$ where $N$ is the length of the tokens sequence. Consider that the codebook provided by VQ-VAE consists of $M$ words, then each entry of the vector $k_i \in \{0,1,\ldots,M\}$ refers to the index of the corresponding word in the codebook. We use a codebook of size $M = 974$, as mentioned in \cite{11}.

Mainly, the output image is constructed through two processes, forward and backward diffusion processes. The forward process gradually corrupts the discrete tokens of the input image via a fixed Markov chain using a transition matrix $Q_t \in \mathbb{R}^{(M+1) \times(M+1)}$. This matrix is constructed using the “Mask-and-replace” strategy \cite{11,35}. Introducing a new special token $\text { [MASK] }$, each token of the image has $M+1$ states so it can be replaced with $\text { [MASK] }$, uniformly replaced with random token or kept unchanged. The newly added token simplifies the reverse process. Further details can be found in \cite{11}. Then, a decoder is trained to estimate the backward diffusion process. The diffusion decoder is trained to denoise and restore the real tokens. The objective of the diffusion decoder is to maximize the conditional probability $q(k|y)$ where $y$ represents the speech embedding vector. The architecture of the decoder consists of 24 blocks. Each block consists of Adaptive Layer Normalization (AdaLN) \cite{30}, full attention, and feed-forward network blocks. We modified the AdaLN to be injected with the speech embedding as a condition to guide the generation process. The AdaLN layer adds the features of the speech with the image features after passing through the full attention.
\subsubsection{Inference}
The input to this stage is only the embedding vector of the spoken caption. Because the image is not an input in the inference stage, there is no forward process. The diffusion decoder input  is completely  random or masked discrete tokens \cite{11,12}. Then, the diffusion decoder tries to restore the real tokens conditioned on the embeddings of the speech input. The diffusion decoder updates the density distribution of all tokens in each step and resamples all tokens according to the new distribution. This strategy prevents accumulating errors produced from predicting the tokens sequentially from left to right and top to down. Finally, we use the VQ-VAE to decode the predicted tokens to generate the output image.

\begin{figure}[!ht]
  \centering
\includegraphics[width=1\linewidth,height=4cm]{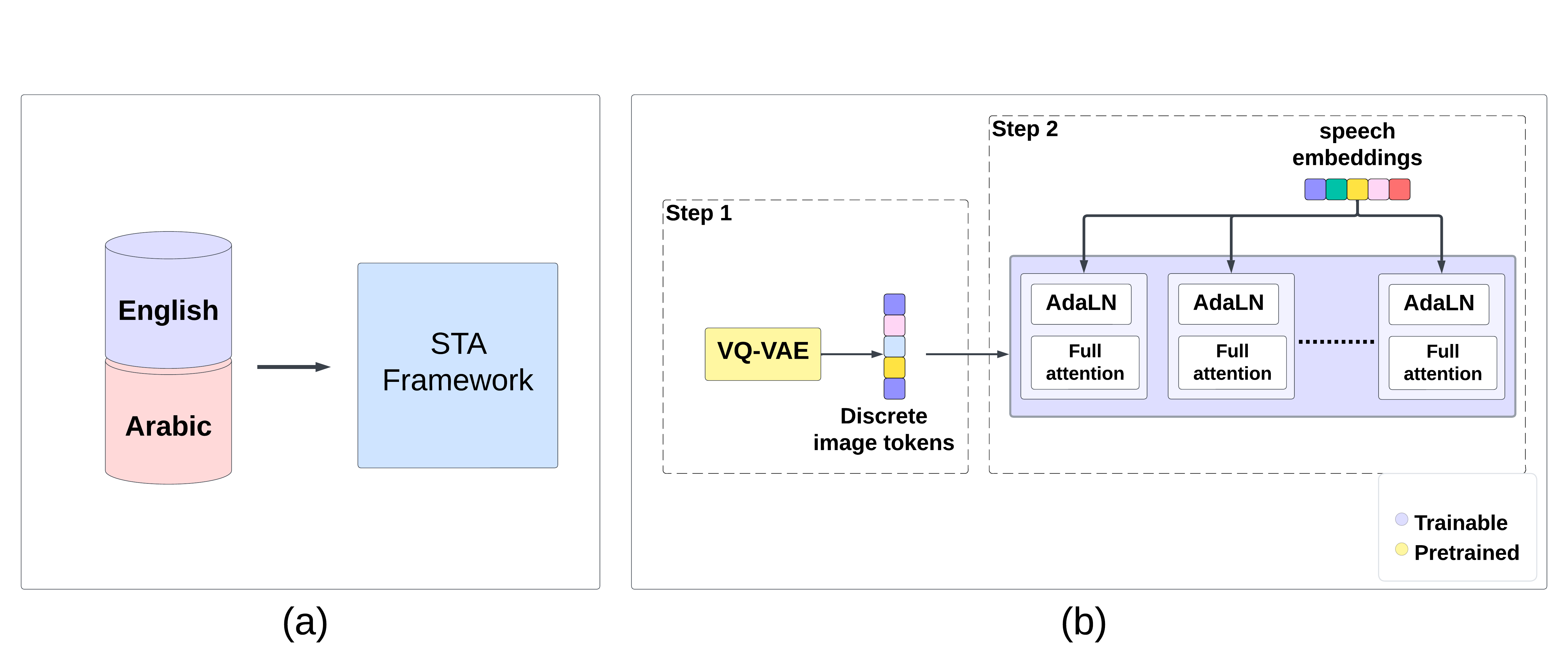}
  \caption{(a) An example of a multilingual dataset. (b) The architecture of VQ-Diffusion model.}
  \label{fig:diffusion}
\end{figure}

\subsection{Multilinguality}
To investigate multiliguality, Our experiments follow a similar approach to M-SpeechCLIP \cite{19}. This approach is independent of the model architecture and could simply be carried out by having a mixed language dataset. Our main goal is to investigate the impact of adding more languages on the performance of the model. As a starting point, Our extended model, Multilingual STA (MSTA), is trained to work with both English and Arabic. As shown in Figure \ref{fig:diffusion}(a), the dataset is a combination of English and Arabic spoken captions, where every English spoken caption has a corresponding Arabic caption.
Our model has been proven to work well in both languages. So, it can possibly be extended to include more languages since English and Arabic belong to different language families.

%% file: sec/4_experiments.tex
\section{Experiments}
\subsection{Training Details}
All experiments are done on Google Colab pro$+$ using  NVIDIA A100 SXM4 40GB.
\textbf{For the speech encoding network training}, we use Adam optimizer with a weight decay of $10^{-6}$, batch size of $64$, and a maximum of $30$ epochs with early stopping if the validation loss doesn't decrease and select the best checkpoint. 
\textbf{For the VQ-Diffusion model training}, we convert $256\times256$ images into $16\times16$ tokens. We set the batch size to $32$, timestamps $T = 100$, and loss weight $\lambda = 0.001$. The network is optimized using AdamW with $\beta_1 = 0.9$ and $\beta_2 = 0.96$. The learning rate is set to $0.00045$ after $5000$ iterations of warmup and the model is trained for $600$ epochs.

\subsection{Datasets}
\subsubsection{CUB-200}
CUB-200 \cite{31} is a dataset of birds images with 200 classes. Each image has 10 English text captions and 10 Arabic text captions \cite{32}. For the task of speech-to-image generation, English-spoken captions were synthesized from text captions using an ASR tool by \cite{2}. We followed the same approach to obtain Arabic-spoken captions. We used the split illustrated in Table \ref{table:datasets}. The classes of the test set do not overlap with the classes of the train and dev sets for a fair comparison with \cite{2}. The train and dev sets include samples of the same classes. The dev set has 15 images randomly selected from each of the 150 classes, the training set has the remaining images.
\input{Tables/datasets}

\subsubsection{Oxford-102}
 Oxford-102 \cite{39} is a dataset of flower images with 102 classes. It contains 8189 images and each image has 10 English text captions. The synthesis process of English-spoken captions and the data split are similar to that of the CUB-200 dataset. 
\subsubsection{Flickr8K}
Flickr8K \cite{20} is a dataset of scene images. Each image has 5 English text captions. Spoken English captions are generated by humans uttering text captions \cite{21}. We translated English text captions into Arabic text captions and synthesized Arabic-spoken captions with an ASR tool. Each image has a total of 10 captions.
\subsection{Evaluation Metrics}
We use Fr\'{e}chet Inception Distance (FID), Inception Score (IS), and Recall@k for evaluating the quality of generated images.
\subsubsection{FID}
FID \cite{22} measures the difference between the Gaussian distributions of the generated images and the ground truth images. We compute FID between the set of images generated from test split captions and the whole set of ground truth images. The lower the FID score, the less the difference between the two distributions is, which means that the model is able to generate images similar to the real ones. To compute FID, image features are extracted using the pre-trained inception model. Each distribution is modeled by its mean \(\mu\) and covariance matrix \(\Sigma\). The FID is then computed as follows
\begin{equation}
\text{FID}\left(x, y\right) = \|\mu_x - \mu_y\|_2^2 + \text{Tr}\left(\Sigma_x + \Sigma_y -2\left(\Sigma_x\Sigma_y\right)^{\frac{1}{2}}\right)
\end{equation}
where $Tr$ is the matrix trace.
\subsubsection{IS}
IS \cite{23} measures the quality and diversity of the generated images. IS is computed as follows
\begin{equation}
\text{IS} = \text{exp}\left(\mathbb{E}_x KL\left(p\left(y|x\right)\| p\left(y\right)\right)\right) 
\end{equation}
where KL is the Kullback-Liebler divergence, \(x\) is a generated image, \(y\) is the label predicted by the Inception-v3 pre-trained on ImageNet \cite{24}. IS has a minimum value of 1 and the maximum value is the number of classes supported by the classification model. The Inception v3 model supports 1000 classes. The higher the IS, the more diverse and meaningful the generated images are.
\subsubsection{Recall@k}
Recall@k \cite{2} is one way to indicate semantic consistency between the generated images and their spoken descriptions. We use the image-image retrieval task between the generated and ground truth images. In image-image retrieval, all ground truth images in the test set are used as queries to retrieve their corresponding synthesized images. Recall@k can then be obtained by computing the percentage of the queries for which at least one matching synthesized image is found among the top-k retrieval results. We use Recall@50 following \cite{2}.

%% file: Tables/datasets.tex
\begin{table}[!ht]
\caption{The splits for CUB-200, Oxford-102, and Flickr8K datasets.}

\begin{tabular}{c|ccc}
\hline
Dataset                     & Train set     & Dev set       & Test set     \\ \hline
\multirow{2}{*}{CUB-200 \cite{31}}    & 6605          & 2250          & 2933         \\
                            & (150 classes) & (150 classes) & (50 classes) \\ \hline
\multirow{2}{*}{Oxford-102 \cite{39}} & 5938          & 1096          & 1155         \\
                            & (82 classes)  & (82 classes)  & (20 classes) \\ \hline
Flickr8K \cite{20}                    & 6000          & 1000          & 1000         \\ \hline
\end{tabular}
\label{table:datasets}
\end{table}

%% file: sec/5_results.tex
\section{Results}
\subsection{Experimental Results on CUB-200 and Oxford-102 English Datasets}\label{subsec:Result_CUB}
We demonstrate the effectiveness of our approach by utilizing synthesized spoken captions as the input to our model instead of textual captions. The synthesized speech data used in our study is of high quality with minimal noise. However, it is worth noting that these speech data are continuous and unaligned, which presents a unique challenge in our research.

CUB-200 and Oxford-102 datasets enable us to make a direct comparison with established techniques for generating images from text. This is because CUB-200 and Oxford-102 are commonly used in tasks related to text-to-image generation, as well as in previous research on direct speech-to-image models.
\input{Tables/CUB}
\input{Tables/oxford}
\subsubsection{Quantitative Results}
In Table \ref{table:cub}, first we compare the results of our model with several state-of-the-art T2IG methods, i.e., StackGANv2, AttnGAN, MirrorGAN, SEGAN, and VQ-Diffusion. As mentioned in \cite{1}, the performance of T2IG can be considered an upper bound for the performance of S2IG frameworks that utilize text as an intermediate stage. Then, we compare our results to the previously released direct speech-to-image generation model.   
It can be noticed from Table \ref{table:cub} and \ref{table:oxford} that our model STA achieves the best results in FID compared to the other models. Subsequently, STA reduces the gap between the performance of text-to-image generation and speech-to-image generation. Since the test set contains non-overlapping classes with the training and test sets, we can conclude that STA generalizes well for new classes of images.
\subsubsection{Qualitative Results}
\begin{figure*}[!ht]
  \centering
  \includegraphics[width=0.65\textwidth]{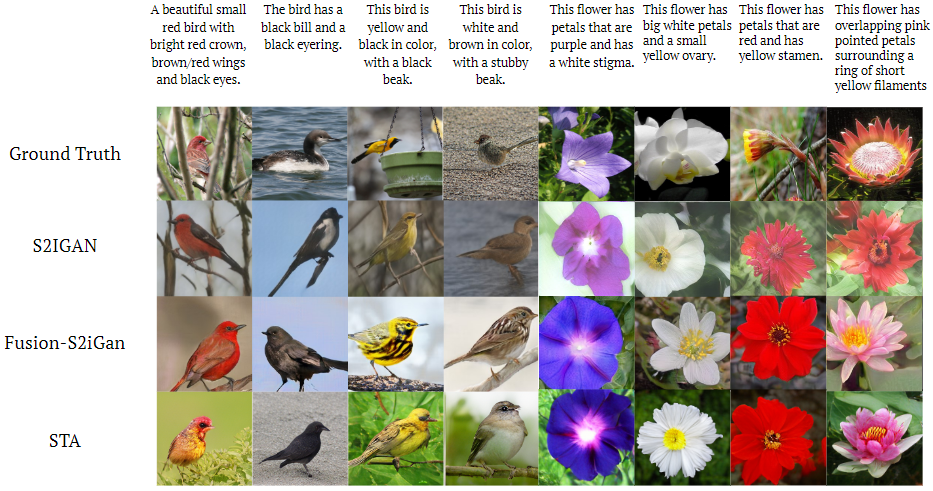}
  \caption{Visual comparison on CUB-200 and Oxford-102 datasets.}
  \label{fig:cub_result}
\end{figure*}

Figure \ref{fig:cub_result} shows 8 selected images generated by S2IGAN and Fusion-S2iGan along with their ground truth images. STA produces impressive high-quality images that capture the semantics of the corresponding spoken caption. The colors of generated images are more vivid, the backgrounds appear to be natural, and the birds are photo-realistic.

\subsection{Experimental Results on Flickr8k English Dataset}\label{subsec:Result_Flickr}
In addition to evaluating our model's performance on a synthesized dataset, we also assessed its effectiveness on the Flickr8k dataset with real speech captions. Synthesized speech is typically more controlled and consistent. In contrast, real speech presents more challenges due to the variability in pronunciation caused by differences in speakers, speaking rate, background noise, and other factors. Therefore, evaluating our model on real speech data is more realistic.
\input{Tables/flickr}

\subsubsection{Quantitative Results}
Table \ref{table:flickr} shows that our model outperforms all the state-of-the-art speech-to-image generation methods with a large margin. Also, our model surpasses AttnGan in the FID metric and has comparable results regarding IS and Recall@50 metrics.
\subsubsection{Qualitative Results} 
\begin{figure*}[!ht]
  \centering
  \includegraphics[width=0.65\textwidth]{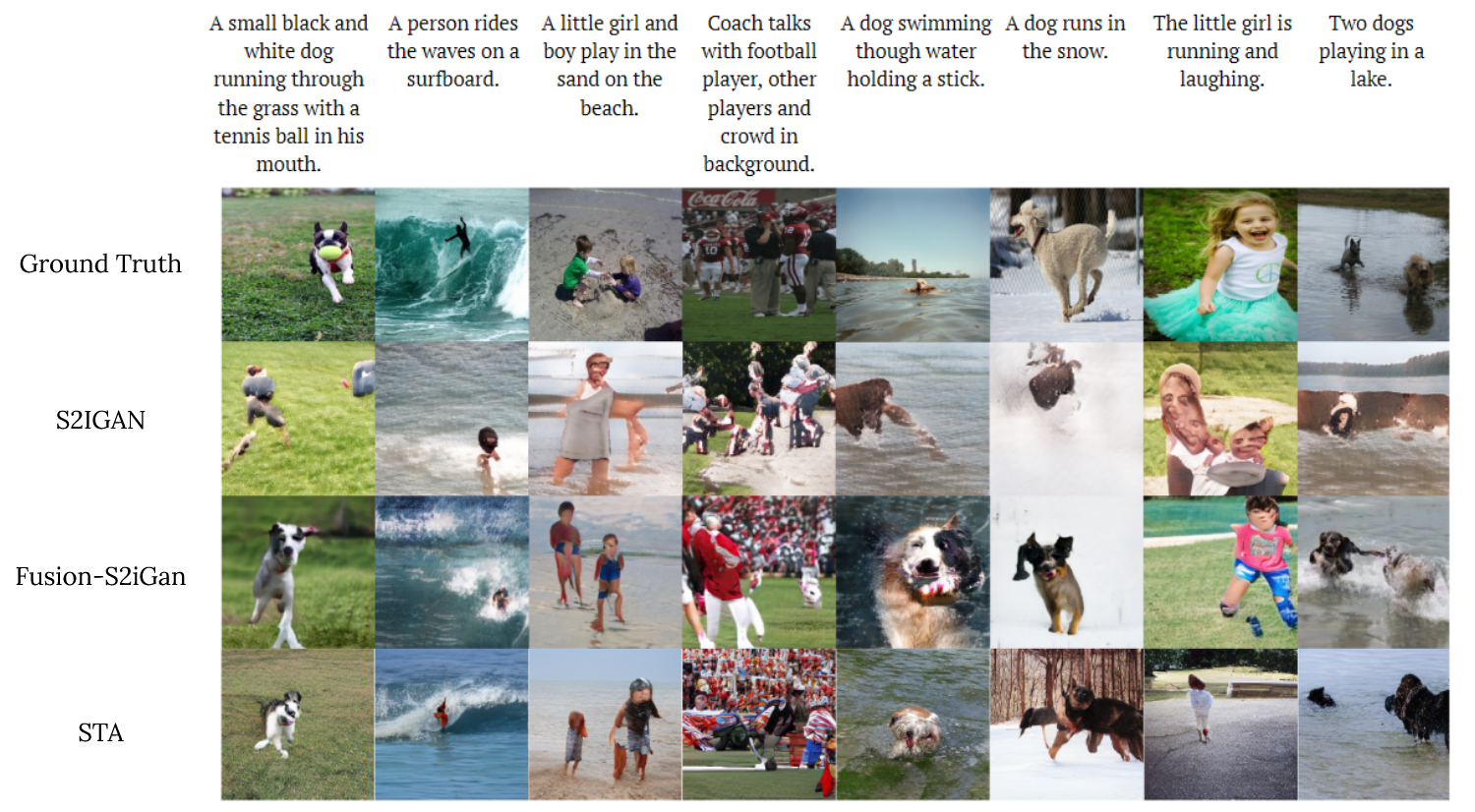}
  \caption{Visual comparison on Flickr8k dataset.}
  \label{fig:flickr_result}
  \vspace{-0.55cm}
\end{figure*}

Flickr8k dataset is highly diverse with a variety of scenes but a limited number of examples per scene. This makes Flickr8k more complex and challenging than datasets with a single object per image, such as CUB-200. Despite its complexity, STA succeeds in synthesizing semantically consistent scenes as shown in Figure \ref{fig:flickr_result}. Images synthesized by STA show more photo-realistic and variant backgrounds than images generated by S2IGAN and Fusion-S2iGan.

\input{Tables/mixed}

\input{grids/multiligual_result}
\subsection{Experimental Results on Multilingual Datasets}
We attempted a novel experiment by aligning the features of two languages in a joint embedding space, as different languages can benefit from each other when learned together. We carried out experiments on both datasets by combining English and Arabic captions.
\subsubsection{Quantitative Results}
We compare the results obtained by our MSTA on the English and Arabic datasets. We also compare the multilingual model MSTA with the performance of the single-language model STA. We have already established in Sections \ref{subsec:Result_CUB} and \ref{subsec:Result_Flickr} that our STA outperformed the previous speech-to-image models. Therefore, we took STA as an upper bound for investigating multilinguality. 
As shown in Table \ref{table:mixed}, The performance of MSTA is similar to that of STA. This indicates that adding another language to our model does not weaken its performance. Moreover, the performance of both languages appears to be approximately the same. 

\subsubsection{Qualitative Results}
We visualize the images produced by MSTA and their corresponding ground-truth images on the CUB-200 and Flickr8k datasets in Figure \ref{fig:compare}.
Comparing images generated from English and Arabic spoken captions, both languages produce images of similar qualities. However, it may be the case that one language can miss some features of the spoken caption, such as colors, while the other language captures those features successfully.


%% file: Tables/CUB.tex
\begin{table}[!ht]
\caption{Performance of STA compared to previous Text-to-Image
and Speech-to-Image models on CUB dataset.}
\centering
    \begin{tabular}{ cccccc }
        \hline
        Dataset&Method&Input&FID ↓&IS ↑\\
        \hline
        \multirow{10}{4em}{CUB-200} & StackGAN-v2 \cite{6} & text & 15.30 & 4.04 ± 0.05 \\ 
        & AttnGAN \cite{7} & text & - & 4.36 ± 0.03\\ 
        & MirrorGAN \cite{8} & text & - &4.56 ± 0.05 \\ 
        & SEGAN \cite{9} & text & - & 4.67 ± 0.04 \\
        & VQ-Diffusion \cite{11} & text & 10.32 & -\\\cmidrule{2-5}
        & Li et al. \cite{1} & speech & 18.37 & 4.09 ± 0.04\\
        & StackGAN-v2 \cite{2} & speech & 18.94 & 4.14 ± 0.04 \\
        & S2IGAN \cite{2} & speech & 14.50 & 4.29 ± 0.04 \\
        & Fusion-S2iGan \cite{15} & speech & 13.09 & \textbf{5.06 ± 0.09} \\
        & STA & speech & \textbf{9.76} & 4.07  ± 0.05 \\
        \hline
    \end{tabular}
    \label{table:cub}
\end{table}




%% file: Tables/oxford.tex
\begin{table}[!ht]
\caption{Performance of STA compared to previous Text-to-Image
and Speech-to-Image models on Oxford dataset.}
\centering
    \begin{tabular}{ cccccc }
        \hline
        Dataset&Method&Input&FID ↓&IS ↑\\
        \hline
        \multirow{7}{4em}{Oxford-102} & StackGAN-v2 \cite{6} & text & 48.68 & 3.26 ± 0.01 \\
        & VQ-Diffusion \cite{11} & text & 14.1 & -
        \\\cmidrule{2-5}
        & Li et al. \cite{1} & speech & 54.76 & 3.23 ± 0.05\\
        & StackGAN-v2 \cite{2} & speech & 54.33 & 3.69 ± 0.08 \\
        & S2IGAN \cite{2} & speech & 48.64 & 3.55 ± 0.04 \\
        & Fusion-S2iGan \cite{15} & speech & 40.08 & \textbf{3.81 ± 0.08} \\
        & STA & speech & \textbf{25.48} & 3.70  ± 0.07 \\
        \hline
    \end{tabular}
    \label{table:oxford}
\end{table}

%% file: Tables/flickr.tex
\begin{table}[!ht]
\caption{Performance of STA compared to previous Text-to-Image and Speech-to-Image models on Flickr8k dataset.}
\begin{center}
\centering
 \resizebox{1.01\linewidth}{!}{
\begin{tabular}{ ccccccc } 
\hline
Dataset&Method&Input&FID ↓&IS ↑& R@50 ↑\\
\hline
\multirow{5}{4em}{Flickr8k} & AttnGAN \cite{7} & text & 84.08 & 12.37 ± 0.41 & \textbf{50.40}\\\cmidrule{2-6}
&StackGAN-v2 \cite{2} & speech & 101.74 & 8.36 ± 0.39 & 16.40\\
&S2IGAN \cite{2} & speech & 93.29 & 8.72 ± 0.34 &16.40\\
&Fusion-S2iGan \cite{15} & speech & 70.80 & 11.70 ± 0.45 & 34.95\\
&STA & speech & \textbf{31.15} & \textbf{12.3 ± 0.6} & 43.46\\
\hline
\end{tabular}
 }
 \end{center}
\label{table:flickr}
\end{table}

%% file: Tables/mixed.tex
\begin{table}[!ht]
\caption{Performance of our multilingual model, MSTA on both mixed datasets.}
\centering
\begin{tabular}{ ccccccc } 
\hline
Dataset&Method&Input&FID ↓&IS ↑& R@50 ↑\\
\hline
\multirow{4}{4em}{Mixed CUB200} & \multicolumn{5}{c}{English}\\\cmidrule{2-6}
& STA & speech & \textbf{9.76} & 4.07 ± 0.05 & -\\
& MSTA & speech & 9.82 & \textbf{4.21 ± 0.06} & -\\\cmidrule{2-6}
&\multicolumn{5}{c}{Arabic}\\\cmidrule{2-6}
& MSTA & speech & 9.87 & 4.15 ±  0.04 & -\\
\hline
\multirow{4}{4em}{Mixed Flickr8k} & \multicolumn{5}{c}{English}\\\cmidrule{2-6}
& STA & speech & \textbf{31.15} &  12.3 ± 0.6 & \textbf{43.46}\\
& MSTA & speech & 31.71 & \textbf{13.06 ± 0.43} & 43.26\\\cmidrule{2-6}
&\multicolumn{5}{c}{Arabic}\\\cmidrule{2-6}
& MSTA & speech & 31.58 & 12.85 ± 0.76 & 41.95\\
\hline
\end{tabular}

\label{table:mixed}
\end{table}

%% file: grids/multiligual_result.tex
\begin{figure*}[hbt!]
\captionsetup[subfloat]{labelformat=empty,font=footnotesize,justification=centering}
\centering
\small
\setlength{\tabcolsep}{3pt}
\begin{tabular}{@{}ccccc@{}}
\raisebox{4\height}{(a)} &
\subfloat[]{\includegraphics[width=0.15\textwidth,height=1.75cm]{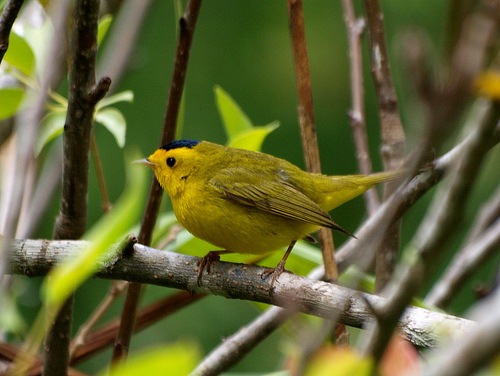}} &
\subfloat[]{\includegraphics[width=0.15\textwidth,height=1.75cm]{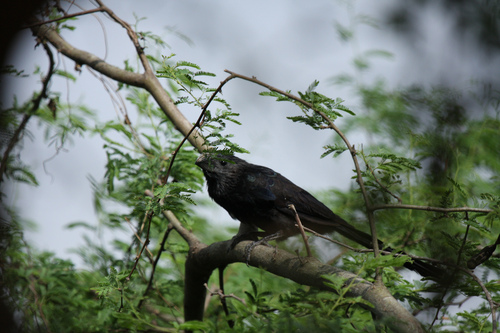}}&
\subfloat[]{\includegraphics[width=0.15\textwidth,height=1.75cm]{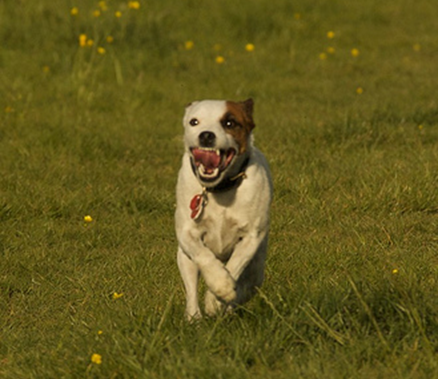}} &
\subfloat[]{\includegraphics[width=0.15\textwidth,height=1.75cm]{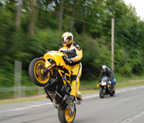}}
\vspace{-0.7cm}
\\
\raisebox{4\height}{(b)} &
\subfloat[\scriptsize This bird is yellow with black on its head and has a very short beak.]{\includegraphics[width=0.15\textwidth,height=1.75cm]{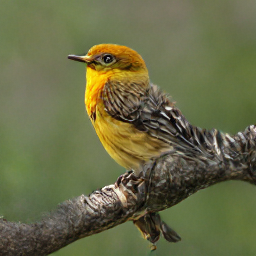}}&
\subfloat[\scriptsize This bird is all black and has a very short beak.]{\includegraphics[width=0.15\textwidth,height=1.75cm]{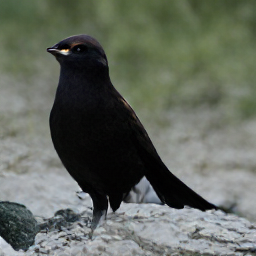}}&
\subfloat[\scriptsize A dog running toward the camera in a field.]{\includegraphics[width=0.15\textwidth,height=1.75cm]{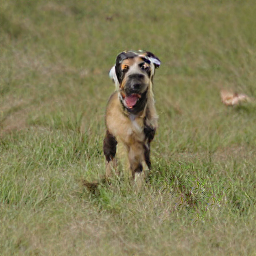}} &
\subfloat[\scriptsize A yellow motorcyclist is popping a wheelie.]{\includegraphics[width=0.15\textwidth,height=1.75cm]{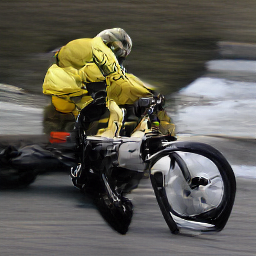}}
\vspace{-0.25cm}
\\
\raisebox{4\height}{(c)} &
\subfloat[ \centering \scriptsize \<هذا الطائر أصفر مع رأس\\ أسود وله منقار قصير للغاية>]{\includegraphics[width=0.15\textwidth,height=1.75cm]{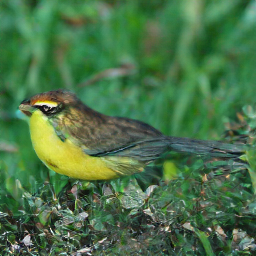}}&
\subfloat[\centering \scriptsize \<هذا الطائر أسود بالكامل \\وله منقار قصير جدًا>]{\includegraphics[width=0.15\textwidth,height=1.75cm]{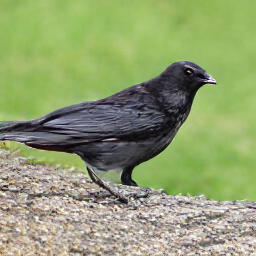}}&
\subfloat[\centering \scriptsize \<كلب يجري باتجاه\\ الكاميرا في حقل>]{\includegraphics[width=0.15\textwidth,height=1.75cm]{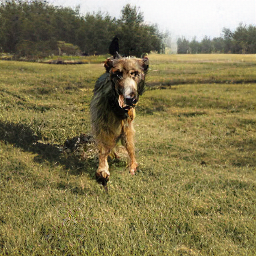}} &
\subfloat[\centering \scriptsize \<يقوم سائق دراجة نارية\\ صفراء بحركة بهلوانية بالدراجة>]{\includegraphics[width=0.15\textwidth,height=1.75cm]{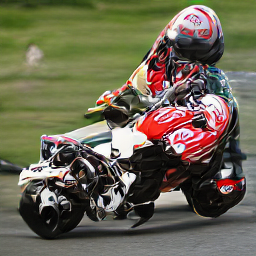}}
\\
\end{tabular}
\caption{Results of MS2IVQ-D model on mixed CUB-200 and mixed Flickr8k datasets. (a) Ground Truth, (b) English captions, and (c) Arabic captions}
\label{fig:compare}
\end{figure*}
\vspace{-0.5cm}

%% file: sec/6_ablation_study.tex
\section{Ablation Study}
In this section, we show the effectiveness of each component in STA. We consider S2IGAN as a base of comparison. For a fair comparison, we trained all models from scratch. For this study, We used Flickr8k dataset to test our framework with real speech which is more challenging.
\input{Tables/ablation_study} 
\input{Tables/ablation_study3}

Firstly, we evaluated the effectiveness of our speech encoding network \cite{16} compared to S2IGAN\cite{2} by replacing S2IGAN's speech encoder with ours. To measure the ability of the model to align the two modalities of speech and images, we performed speech-image retrieval to evaluate the performance of the speech encoding network. As shown in Table \ref{table:ablation3}, the speech encoding network of STA outperformed that of S2IGAN.

Secondly, we evaluate the performance of VQ-diffusion. We used the embeddings extracted from our speech encoding network as a condition for training the stacked GAN of S2IGAN \cite{2} and compare the final results with our framework. As shown in Table \ref{table:ablation}, the VQ-Diffusion is more effective than the stacked GAN \cite{2} in generating diverse and complex scenes.

%% file: Tables/ablation_study.tex
\begin{table}[!ht]
\caption{Results with replacing different parts of S2IGAN \cite{2} compared to S2IGAN \cite{2} on Flickr dataset. Results are reported at its best epoch.}
\centering
\begin{tabular}{cccc}
\hline
Method & FID ↓ & IS ↑  & R@50 ↑   \\
\hline
S2IGAN\cite{2} & 93.29    & 8.72 ± 0.34 & 16.40    \\
\begin{tabular}{@{}c@{}} STA's speech encoding  \\ network + GAN\cite{2} \end{tabular}   & 68.78    & 9.11 ± 0.4  & 20.02     \\
STA & \textbf{31.15} & \textbf{12.3 ± 0.6} & \textbf{43.46} \\
\hline
\end{tabular} 
\label{table:ablation}
\end{table}

%% file: Tables/ablation_study3.tex
\begin{table}[!ht]
\caption{Comparison of retrieval scores of different speech encoding networks on Flicker8k dataset.}
\centering
\begin{tabular}{ccccccc}
\hline
        & \multicolumn{3}{c}{Speech → Image} & \multicolumn{3}{c}{Image → Speech}\\

\cmidrule(lr){2-4}\cmidrule(lr){5-7} 

Method  & R@1 & R@5 & R@10 & R@1 & R@5 & R@10    \\
\hline
S2IGAN \cite{2} 
        & 6.39  & 18.76 &29.48 & 5.88 & 20.21 & 30.21  \\

STA
        & \textbf{40.89} & \textbf{74.37} & \textbf{84.99} & \textbf{52.78} & \textbf{85.46} & \textbf{92.47} \\
\hline
\end{tabular} 
\label{table:ablation3}
 \vspace{-0.5cm}
\end{table}


%% file: sec/7_conclution.tex
\section{Conclusion}
\normalsize
In this work, we propose STA, a speech-to-image synthesis framework. Speech, as input, is more complex than text because speech can be noisy or contain irrelevant information that can mislead the model during training. However, with a well-trained speech encoder, the performance of the speech-to-image task approaches the text-to-image task. Moreover, the use of diffusion models helps improve the quality of the generated images. Our proposed method reduces the gap between speech and text models for conditioned image generation and is not restricted to two languages. Multilinguality is supported by mixing spoken captions of multiple languages for the same dataset. Using STA for real-life applications requires training on large datasets such as LAION \cite{25}.